\documentclass[12pt,aps,prb,preprint,showpacs,showkeys]{revtex4}
\usepackage{amsmath}
\usepackage{amsfonts}
\usepackage{amssymb}
\usepackage{bm}
\usepackage[pdftex]{graphicx}

\newcommand{\kw}{$kW/cm^2$\ }
\newcommand{\cm}{$cm^{-1}$\ }
\newcommand{\nss}{NSs}
\begin{document}

\title{Thermal effects on electron-phonon interaction in silicon nanostructures}

\author{Rajesh Kumar}
\email{rajesh2@ualberta.ca}
\thanks{Present Address: National Institute for Nanotechnology (NINT), University of Alberta, Edmonton T6G 2M9, CANADA}
\author{Vivek Kumar}
\author{A. K. Shukla}
\email{akshukla@physics.iitd.ernet.in}

\affiliation{Department of Physics, Indian Institute of Technology
Delhi, New Delhi 110016, India}

\begin{abstract}
Raman spectra from silicon nanostructures, recorded using excitation
laser power density of 1.0 \kw, is employed here to reveal the
dominance of thermal effects at temperatures higher than the room
temperature. Room temperature Raman spectrum shows only phonon
confinement and Fano effects. Raman spectra recorded at higher
temperatures show increase in FWHM and decrease in asymmetry ratio
with respect to its room temperature counterpart. Experimental Raman
scattering data are analyzed successfully using theoretical Raman
line-shape generated by incorporating the temperature dependence of
phonon dispersion relation. Experimental and theoretical temperature
dependent Raman spectra are in good agreement. Although quantum
confinement and Fano effects persists, heating effects start
dominating at higher temperatures than room tempaerature.
\end{abstract}
\date{\today}

\maketitle

Raman scattering from all forms of solids (single crystal, poly-and
nanocrystalline) gives many information about its vibrational
properties and the crystallanity. Quantum confinement effect in
semiconductor nanostructures(\nss) can also be investigated very
effectively by Raman spectroscopy by close analysis of the Raman
line-shape. A symmetric Raman spectrum with very narrow line width
is observed from crystalline materials as a signature of zone center
optic phonon mode having Lorentzian line-shape. Any deviation in
Raman line-shape symmetry, frequency or line width could be due to
various reasons associated with quantum confinement effect, heating
effect, electron-phonon interactions or any combination of these.
For silicon (Si), a 4 \cm broad symmetric Raman line-shape centered
at 520.5 \cm is observed at room temperature. Temperature dependence
of Raman line-shape from crystalline Si (c-Si) has been studied by
many authors \cite{1,2,3,4}. The Lorentzian Raman line-shape from
c-Si becomes wider and shifts to lower wavenumber side at elevated
temperature {\cite{3}}. As a result of increase in the temperature
the stokes and antistokes intensity ratio also decreases, which in
turn is used to calculate the sample temperature {\cite{4,5}}.
Asymmetric Raman line-shapes are observed from heavily doped Si due
to Fano interaction \cite{6}. Fano interaction takes place as a
consequence of interference between discrete phonons and continuum
of electronic states (available as a consequence of heavy doping) in
heavily doped Si {\cite{7,8}}. Asymmetric Raman line-shapes can also
be observed from nanostructures (NSs) {\cite{9,10,11}}. In Si NSs,
the phonons are confined within the small NSs causing the breakdown
of $\overrightarrow{k} = 0$ Raman selection rule and all optic
phonons take part in the Raman scattering {\cite{12,13}}. As a
result Raman line-shape becomes asymmetric and downshifted for Si
NSs.

Attempts have been taken to investigate a combination of different
effects that may change Raman line-shape. Recently, a combined
effect of photo-excited Fano interaction and phonon confinement is
also reported for Si NSs {\cite{14,15}}. Temperature dependent Raman
scattering studies for Si NSs for temperatures above 300 K have been
reported previously {\cite{16,17,18,19,20,21}} by combining heating
and confinement effects. Many authors {\cite{17,18,19,20,21}}
explain the Raman line-shape at elevated temperature by simply
considering the temperature dependence of Raman peak position and
FWHM as formulated by Balkanski et al {\cite{4}}. He used the higher
order anharmonicity in light scattering by optical phonons to
incorporate the effect of temperature on zone center optic phonons
only. Narsimhan and Vanderbilt \cite{22} have studied the
phonon-phonon interaction in Si to calculate the contribution from
anharmonicity to phonon lifetime and frequency shifts. They have
shown that $\Gamma$ and L point optic phonons show different
temperature dependence from each other. This means that not only the
zone center phonons but the whole phonon dispersion relation change
as a function of temperature. Thus, one may expect that temperature
dependence of Raman line-shape from Si NSs (where all phonons
participate in the Raman scattering) should be calculated by
considering the temperature dependence of whole phonon dispersion
relation and not by simple considering the peak shift. Same
consideration should be taken care of when there is a possibility of
Fano interaction as well.

In this Report, we  present temperature dependence of photo-excited
Fano interaction in Si NSs prepared by laser induced etching (LIE)
technique \cite {23}. Raman spectra are recorded using excitation
laser power density of 1.0 \kw at four sample temperatures in the
range 300 K - 600 K. Room temperature Raman spectra for excitation
laser power density of 1.0 \kw shows a combined effect of phonon
confinement and photo-excited Fano interaction. Experimental Raman
spectra recorded at higher temperature show effect of temperature as
well in terms of changes in the Raman line-shape. Temperature
dependent changes in Raman line-shape have been explained
theoretically by considering the temperature dependence of whole
phonon dispersion relation as suggested by Narsimhan and Vanderbilt
\cite{22}

The Si NSs sample under investigation is fabricated by the LIE
technique \cite {23}. The LIE is done by immersing a Si wafer 48\%
HF acid and then focusing a 500 mW argon-ion laser beam (E$_{ex}$ =
2.41 eV) for 45 minutes. Raman scattering was excited using photon
energy 2.41 eV of the argon-ion laser with laser power density of
1.0 \kw. The reason for choosing low laser power density is to avoid
heating of the sample during Raman recording. Sample was heated
using the heating arrangement on the sample holder to study the
temperature dependent Raman scattering. Raman spectra were recorded
by employing a SPEX-1403 doublemonochromator with HAMAMATSU (R943-2)
photomultiplier tube arrangement and an argon ion laser (COHERENT,
INNOVA 90).

Figure 1 shows the high resolution atomic force microscope (AFM)
image of the Si NSs prepared by LIE technique. The shown AFM image
is taken from the inside of the pore walls in the sample \cite{24}.
The AFM image shows the formation of Si NSs having sizes in the
range of a few nanometers with very narrow size distribution.
Quantum confinement is expected within these Si NSs because the
sizes are comparable to the Bohr exciton radius (5 nm) of Si
\cite{25}. Raman experiments are done to investigate any possibility
of phonon confinement effect in Si NSs. Figure 2(a) shows the room
temperature Raman spectrum from Si NSs recorded using excitation
laser power density of 1 \kw. The asymmetric Raman spectrum in Fig.
2(a) is red-shifted and broader in comparison to its c-Si
counterpart. Asymmetrical broadening and red-shift is attributed to
a combined effect of quantum confinement and photo-excited Fano
interaction \cite {13,14}. The Fano interaction takes place between
the discrete phonon Raman scattering and quasi-continuum electronic
Raman scattering \cite {26}. Presence of quasi-continuum of
electronic states available for electronic Raman scattering have
been investigated using photoluminescence spectroscopy in our
previous reports\cite {26}.

To see the effect of heating on Fano interaction, Raman spectra from
the Si NSs have been recorded  at different temperatures using
excitation laser power density of 1.0 \kw as displayed in Figs. 2(b)
- 2(d). Sample temperature is estimated from the stokes and
antistokes Raman intensity ratio \cite {4, 5}. Raman spectra at
higher temperatures in Figs. 2(b) - 2(d) reveal that the red-shift
and FWHM increase with increasing sample temperature. We attribute
the changes in Raman features to a combined effect of quantum
confinement effect, Photo-excited Fano interaction and heating
effect. To theoretically fit the experimentally observed Raman data
in Fig. 2, we have modified the Raman line-shape from Si NSs by
including the temperature dependence of phonon dispersion relation.
The modified temperature dependent Raman line-shape from Si NSs can
be written as follows to incorporate the temperature dependent Fano
interaction:
\begin{equation}\label{eq. 1}
    I(\omega,T) \propto {\int_{L_1}^{L_2} N(L) \left[ {\int_0^1 \left\{\frac{(\varepsilon+q)^2}{1+\varepsilon^2}\right\} e^{\frac{-k^2 L^2}{4 a^2}}} d^2k\right]dL}
\end{equation}
where, $\varepsilon =\frac{\omega - \omega(k,T)}{\gamma/2}$ and
`$q$' is Fano asymmetry parameter. The `$k$' is the phonon wave
vector. The $\gamma$, L and `$a$' are the line width, crystallite
size and lattice constant respectively. The `$N(L)$' is a Gaussian
function of the form $N(L)\cong e^{-\left\{
\frac{L-L_0}{\sigma}\right\}^2}$ included to account for the size
distribution of the Si NSs. The $L_0$, $\sigma$, $L_1$ and $L_2$ are
the mean crystallite size, the standard deviation of the size
distribution, the minimum and the maximum confinement dimensions
respectively. The $\omega(k,T)$ is the temperature dependent phonon
dispersion relation of the optic phonons of c-Si given by
$\omega(k,T)=\sqrt{A(T)+B(T) cos\left( \frac{\pi k}{2}\right)}$.
Temperature dependent parameters, $A(T)$ and $B(T)$ have been taken
from the analysis done by Narsimhan and Vanderbilt \cite {22}. They
have calculated the downshift in `$\Gamma$' and `L' point optic
phonons as a function of temperature. Thus, the value of `$\Gamma$'
and `L' point phonon frequency can be known at a given temperature.
The `L' point phonon frequency is utilized to find the value of $A$
by using $k = 1$ in the phonon dispersion relation written above and
$B$ can be calculated by utilizing `$\Gamma$'  point ($k = 0$)
phonon frequency value. The values of $A(T)$ and $B(T)$, calculated
in this way for Raman fitting, are summarized in Table \ref
{table:1}. At room temperature (300 K), the Eq. (1) reduces to the
Raman line-shape having only quantum confinement effect and
photo-excited Fano interaction as used in our earlier papers \cite
{14,15}. The room temperature Raman data in Fig. 2(a) is best fitted
for fitting parameters $L_0$, $L_1$, $L_2$ and $\sigma$ of 4 nm, 3
nm, 5 nm and 2 nm respectively. Another fitting  parameter in Eq.(1)
is Fano asymmetry parameter `$q$'. The value of $|q|$ = 11 is
obtained by fitting experimental data in Fig. 2(a) with Eq. (1). The
value of `$q$' was kept constant while fitting the Raman spectra in
Figs. 2(b)-2(d) because thermally generated carriers could be
treated to be negligible for Fano interaction. Theoretically fitted
Raman line-shapes generated using Eq. (1) are shown by continuous
lines in Fig. 2. It is evident from theoretical Raman fitting of
experimental Raman data that room temperature Raman scattering has
contribution only from phonon confinement effect and photo-excited
Fano interaction. Raman spectra at higher temperatures have
additional contribution from heating effect in addition to the
phonon confinement and Fano effects. As a result of heating, the
FWHM and phonon softening increases in Fig. 2 but asymmetry ratio
(asymmetry ratio is defined as $\frac{\gamma_l}{\gamma_h}$ , where
$\gamma_l$ and $\gamma_h$ are the half widths on the lower and
higher side of the Raman peak position) decreases. Asymmetry ratios
and FWHMs of Raman line-shapes in Fig. 2 are compared in Fig. 3.
When sample temperature is increased, Raman half widths increase
equally on both the sides of the peak position. As a result,
asymmetry ratio ($\frac{\gamma_l}{\gamma_h}$) decreases as heating
effect dominates over the phonon confinement and Fano effects, which
is fixed in our sample. In other words, phonon confinement and Fano
effects are dominated by heating effects.

   Temperature dependent experimental Raman data
from Si NSs in Fig. 2 show a good theoretical fitting using Eq. (1).
This means that the temperature dependence of Raman scattering
should be incorporated in theoretical Raman line-shape by
considering the temperature dependence of the whole phonon
dispersion relation as done in our case. This type of analysis is
necessary because the thermal effect is different for $\Gamma$ and L
point phonons as reported by Narsimhan and Vanderbilt \cite {22}. As
a result, $\Gamma$ and L points on optic branch downshift by
different amounts at higher temperatures. This fact could be taken
into account only by taking the temperature dependence of whole
phonon dispersion relation. This analysis will be extremely
important in understanding any non-linear  Fano interaction in
nanoscale materials \cite {27}.

In summary, temperature related effects are found to dominate over
the phonon confinement and Fano interaction in Si NSs in Raman
spectra recorded at higher temperatures than room temperature. The
room temperature Raman spectrum recorded using laser power density
of 1 \kw show contributions from photo-excited Fano interaction and
phonon confinement effect. Increase in FWHM and decrease in
asymmetry ratio in Raman line-shape were observed as a result of
increasing temperature. Theoretical analyses of Raman spectra are
done by fitting the experimental Raman data with theoretical Raman
line-shape obtained by appropriately considering the phonon
confinement effect and Fano interaction. Temperature dependence is
incorporated in theoretical Raman line-shapes by considering the
temperature dependence of whole phonon dispersion relation.
Temperature dependent phonon dispersion relation is obtained by
considering different downshifts in $\Gamma$ and L point phonons in
phonon dispersion curve at a given temperature. A good agreement
between experimental and theoretical Raman results reveals that it
is more appropriate to incorporate temperature dependence of full
phonon dispersion relation to explain the temperature dependent
Raman spectra from Si NSs.

Authors are thankful to Prof. V.D. Vankar for many useful
discussions. Authors acknowledge the financial support from the
Department of Science and Technology, Govt. of India under the
project ``Linear and nonlinear optical properties of
semiconductor/metal nanoparticles for optical/electronic devices".
One of the authors (R.K.) acknowledges financial assistance from
National Research Council (NRC). NINT is operated as a partnership
between the NRC and the University of Alberta and is jointly funded
by the Govt. of Canada, the Govt. of Alberta and University of
Alberta. One of the authors (V.K.) acknowledges the financial
support from University Grants Commission (UGC), India. Technical
support from Mr. N.C. Nautiyal is also acknowledged.

\newpage
\begin{table}
  \centering
  \begin{tabular}{|c|c|c|c|}
    \hline
    Temperature (K) & A($cm^{-2}$) & B($cm^{-2}$) & $\gamma$(\cm) \\
    \hline
    300 & 171400 & 100000 & 4.0 \\
    400 & 170417 & 99011.4 & 4.95 \\
    500 & 169223.6 & 98717.2 & 6.36 \\
    600 & 167819.77 & 98369.25 & 7.93 \\
    \hline
  \end{tabular}
    \caption{Temperature dependent parameters used in Eq.(1)}\label{table:1}
\end{table}

\newpage
\begin{figure}[t!]
\includegraphics[width=10.0cm]{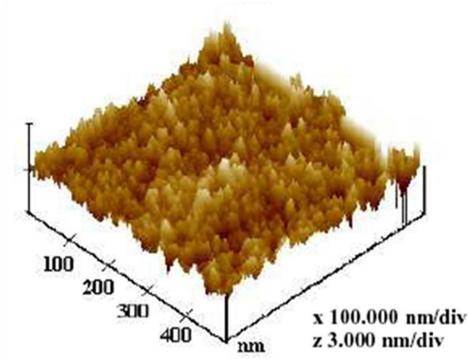}
\caption{(\textbf{Color online}) AFM image of silicon nanostructures
prepared by laser induced etching method.} \label{fig.1}
\end{figure}

\newpage

\begin{figure}[t!]
\includegraphics[width=10.0cm]{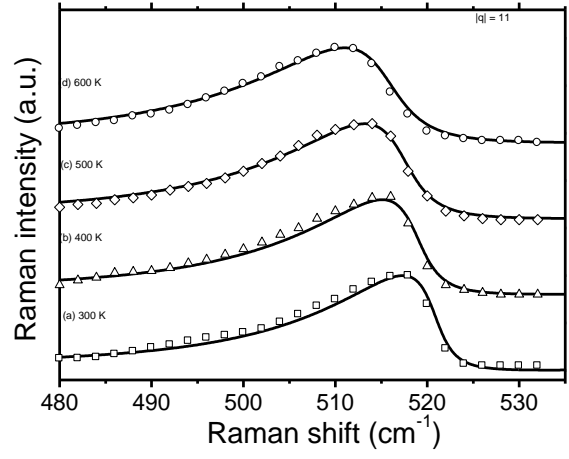}
\caption{Raman spectra from silicon nanostructures at different
sample temperatures from 300 K to 600 K recorded at excitation laser
power density of 1.0 \kw. Discrete points are experimentally
recorded data and continuous lines are theoretical fitting using Eq.
(1) by considering the temperature effect on  photo-excited Fano
effect.} \label{fig.2}
\end{figure}

\newpage

\begin{figure}[t!]
\includegraphics[width=12.0cm]{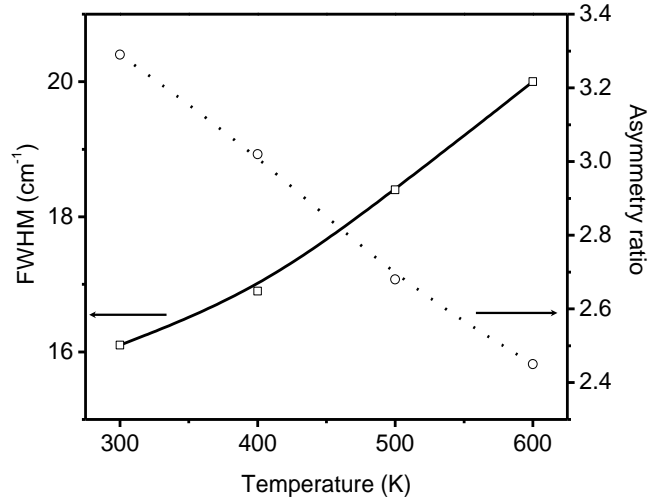}
\caption{Temperature dependent variation of asymmetry ratio and FWHM
of Raman spectra from silicon nanostructures shown in Fig.2.}
\label{fig.3}
\end{figure}

\end{document}